\documentstyle[psfig]{mn}
\def\be{\begin{equation}} \def\en{\end{equation}}

\begin{document}

\title{A  new multidimensional  AMR  Hydro+Gravity Cosmological  code}
\author[Vicent  Quilis]{Vicent Quilis  \\  Departament d'Astronomia  i
Astrof\'{\i}sica,  Universitat   de  Val\`encia,  46100   -  Burjassot
(Valencia), Spain\\ e-mail: vicent.quilis@uv.es}

\maketitle
 
\begin{abstract}
A new cosmological multidimensional hydrodynamic and N-body code based
on an  Adaptive Mesh Refinement  scheme is described and  tested.  The
hydro part  is based  on modern {\it  high-resolution shock-capturing}
techniques,  whereas N-body  approach is  based on  the  Particle Mesh
method.   The code  has  been specifically  designed for  cosmological
applications.  Tests  including shocks, strong  gradients, and gravity
have  been considered.   A cosmological  test based  on  Santa Barbara
cluster is also  presented.  The usefulness of the  code is discussed.
In particular,  this powerful  tool is expected  to be  appropriate to
describe  the  evolution  of  the  hot gas  component  located  inside
asymmetric cosmological structures.
\end{abstract}
 

\section{Introduction}
 
Numerical simulations  of structure  formation are essential  tools in
theoretical  cosmology. Their  main role,  in addition  to  many other
uses,  has  been to  test  the viability  of  the  different models  of
structure formation  -- e.g.  variants of the  cold dark  matter (CDM)
model -- by evolving initial conditions using basic physical laws.

Historically,  the use  of cosmological  simulations started  in 1960s
(Aarseth  1963) and  1970s (e.g.   Peebles 1970,  White  1976).  These
calculations   were  N-body,   collisionless,  simulations   with  few
particles.   In  early  1980s,  the N-body  techniques  and  computers
advanced   to   the  extent   that   simulations   could  be   applied
systematically as a scientific tool ( see Hockney \& Eastwood 1988 for
a good  review of  N-body techniques) and  their use led  to important
results such  as the  development of the  CDM paradigm  (Peebles 1982,
Blumenthal et al.  1984, Davies  et al. 1985).  N-body techniques have
continued to  be developed and  refined, with examples  including AP3M
(Couchman 1991), ART code  (Kravtsov, Klypin \& Khokhlov 1997), GADGET
code (Springel, Yoshida \& White 2001), MLAPM (Knebe, Andrew \& Binney
2001) or more recently the code presented by Bode \& Ostriker (2003).

The physics of the gaseous, baryonic, component of the Universe is far
more complicated to model than the formation of structures in the dark
matter due  to gravitational instability.   Understanding the behaviour
of baryons is crucial for a complete theory of the formation of cosmic
structures, particularly on galactic and cluster scales.  Any realistic
simulation which sets  out to explain the growth  of structures in the
Universe  must therefore  contain  a hydrodynamical  treatment of  the
evolution of the baryonic  fluid.  Pioneering simulations using Smooth
Particle  Hydrodynamics (SPH)  techniques  were first  carried out  by
Gingold  \&  Monaghan  (1977);  early  cosmological  simulations  that
followed baryons  and dark matter  include those by Evrard  (1988) and
Hernquist \& Katz (1989).

The integration of the hydrodynamics equations governing the evolution
of  gas can be  done using  different techniques.   The adoption  of a
particular technique, with its  associated benefits and drawbacks, has
a direct consequence  on the outcome of the  simulation. The numerical
techniques used  to model  the evolution of  the baryons can  be split
into two general classes: Lagrangian and Eulerian.

The most popular of the Lagrangian scheme is SPH(Lucy 1977, Gingold \&
Monaghan   1977).   Other   schemes   which  could   be   defined   as
quasi-Lagrangian  include those  described  by Gnedin  (1995) and  Pen
(1995). SPH techniques  have made possible huge advances  in the field
of  numerical  Cosmology  in  the  recent past.   Relatively  easy  to
implement and with a low  computational cost, SPH techniques have a huge
dynamical range because of  their Lagrangian nature.  This feature has
made them  particularly successful  in simulations of  cosmic structure
formation. However, the SPH  technique also has significant drawbacks,
including (i) 
an  approximate treatment and  description of  shock waves  and strong
gradients, (ii)  a poor  description of low  density regions,  (iii) a
requirement to the  use of numerical artifacts such  as the artificial
viscosity and  (iv) the possible violation  of conservation properties
(see Okamoto et al.  (2003) for a recent example of unphysical results
in simulations due to the SPH nature).

Eulerian schemes present an alternative to Lagrangian schemes.  Within
this board class of techniques, the ones based on Riemann solvers have
been particularly successful (e.g. Ryu et al. 1993; Quilis, Ib\'a\~nez
\& S\'aez 1994;  Bryan, Norman, Stone, Cen \&  Ostriker 1995; Gheller,
Pantano \&  Moscardini 1998, and others).  These  numerical schemes are
written in conservative form,  which ensures excellent conservation of
physical   quantities.   Shock   waves,  discontinuities   and  strong
gradients are sharply resolved in typically one or two cells.  The use
of Riemann solvers removes the  need to invoke artificial viscosity to
integrate equations  with discontinuities.  Although  these properties
are needed in order to build a robust hydrodynamical scheme, precisely
due to their  Eulerian character -- fix numerical  grids are needed to
integrate the hydrodynamical equations -- these techniques are limited
by poor spatial resolution.   In order to achieve adequate resolution,
dense  numerical  grids  are   needed  which  quickly  drives  up  the
computational cost.  Even with  the best computers available nowadays,
a simple  Eulerian approach cannot compete with  a Lagrangian approach
in cosmological  applications, which demand a  good spatial resolution
and a huge dynamical range.

In this paper, we present a  new numerical code which combines the best
features of the Eulerian and Lagrangian approaches.  The basic idea is
to improve the numerical resolution  by implementing a scheme known as
Adaptive  Mesh  Refinement (AMR)  described  originally  by Berger  \&
Oliger (1984) and  Berger \& Colella (1989).  In order  to do this, we
use  an Eulerian  approach as  the ones  described above,  but gaining
resolution -- both spatial and temporal -- by selectively refining the
original computational grid. The result is a hierarchy of nested grids
which naturally behaves as a  Lagrangian scheme (the grids are refined
only where the calculation requires it) and each one of these grids is
treated as an independent computational domain by the Eulerian scheme.

AMR  schemes  have proved  to  be  extremely  powerful in  many  fluid
dynamics applications.  In  cosmology, recent AMR implementations have
been designed by  Bryan \& Norman (1997), Kravtsov,  Klypin \& Hoffman
(2002) and Teyssier  (2002).  These codes share  the basic ingredients
with  the one  described  in  this paper,  although  there are  slight
differences depending on each particular implementation.

The  central  tenet   of  the  AMR  scheme,  the   refinement  of  the
computational  grid  wherever  better  resolution is  needed,  can  be
exploited to incorporate  an N-body scheme to follow  the evolution of
dark  matter.   The Particle-Mesh(PM)  method  is  ideally suited  for
grating onto an hydro-AMR code.  In  practise, a PM scheme is used for
each nested  grid, with  progressively higher spatial  resolution when
the cell size gets smaller.  This implementation of the AMR-PM has the
advantage of avoiding the problem  of setting a softening parameter in
the gravitational force law, as this parameter is naturally determined
by the  cell size. Several  implementations of this approach  for dark
matter  only have  been  presented in  the  literature with  different
degrees  of success  (Villumsen 1989,  Jessop et  al.   1994; Splinter
1996; and Kravtsov, Klypin \& Khokhlov 1997).

In the present paper, we describe a new coupled hydrodynamical 
and N-body code
for cosmological applications --  called MASCLET (Mesh Adaptive Scheme
for CosmologicaL structurE evoluTion) --  based on an AMR scheme.  The
basic hydrodynamical solver is based on the Piecewise Parabolic Method
(Colella \& Woodward 1984) whereas the N-body method used is a classic
PM according  to Hockney \&  Eastwood (1988). The scheme  presented in
this  paper  is  able  to  refine  and  unrefine  grids  automatically
according with a set of  parameters which can be chosen depending upon
the application.   The code is  written in Fortran  90 and there  is a
parallel version using OpenMP standard directives.

The following  section is devoted to present  the equations describing
the evolution  of gaseous  and dark matter  components, and  the main
features  of the  basic hydrodynamical  and N-body  schemes for  a fix
grid. Sec. 3  describes the AMR strategy and how  it is implemented in
the code. In Sec. 4,  we present several numerical tests including the
so-called Santa Barbara cluster (Frenk et al. 1999). Finally, our main
conclusions are summarized in Sec. 5. 
  
\section{Equations and the basic numerical procedure}
 
In  this Section  we  write  down the  basic  equations governing  the
evolution of cosmological inhomogeneities  as a {\it hyperbolic system
of  conservation laws}. The  mathematical properties  of this  kind of
system and the numerical  algorithms specifically designed for solving
it have been well studied in the literature (LeVeque 1992, Toro 1997).
 
\subsection{Gas dynamics} 
 
\subsubsection{Evolution equations in conservation form}

For spatial  scales where  relativistic corrections are  not required,
cosmological  inhomogeneities   evolve  according  to   the  following
equations (Peebles 1980):
\begin{equation}
\frac{\partial   \delta}{\partial  t}  +   \frac{1}{a}\nabla\cdot  (1+
\delta) {\bf v} = 0
\label{hydro1}
\end{equation}
\begin{equation}
\frac{\partial  {\bf v}}{\partial t}  + \frac{1}{a}  ({\bf v}\,\cdot\,
\nabla){\bf v} +  H{\bf v} = - \frac{1}{\rho  a}\nabla p - \frac{1}{a}
\nabla \phi
\label{hydro2}
\end{equation}
\begin{equation}
\frac{\partial E}{\partial  t} + {1\over  a} \nabla\cdot[(E +  p) {\bf
v}] = -3H(E+p) -H\rho {\bf v}^2 - \frac{ \rho {\bf v}}{a} \nabla \phi
\label{hydro3}
\end{equation}
\begin{equation}
\nabla^2\phi = \frac{3}{2} H^2 a^2 \delta
\label{poisson1}
\end{equation}
where ${\bf  x}$, ${\bf v}=a(t)\frac{d{\bf x}}{dt}=  (v_x, v_y, v_z)$,
and $\phi(t,{\bf x})$ are, respectively, the Eulerian coordinates, the
peculiar velocity, and the peculiar Newtonian gravitational potential.
The total energy  density, $E= \rho \epsilon + {1\over  2} \rho v^2$ ,
is  defined as  the addition  of the  thermal  energy, $\rho\epsilon$,
where  $\epsilon$ is  the specific  internal energy,  and  the kinetic
energy  (where  $v^2  =  v_x^2  + v_y^2  +  v_z^2$).   The  background
parameters   are   the   scale   factor,  $a$,   background   density,
$\rho_{_{B}}$, and  the Hubble constant,  $H$. The density  contrast ,
$\delta$,  is  defined  as  $\delta=\rho/\rho_{_{B}} -  1$.   Pressure
gradients  and  gravitational  forces  are  the  responsible  for  the
evolution.
  
Poisson's equation  (\ref{poisson1}) is an elliptic  equation, and its
solution depends on  the boundary conditions. This equation  has to be
solved in conjunction with Eqs (\ref{hydro1}--\ref{hydro3}) to compute
the    source   term    $\nabla   \phi$    which   appears    in   Eqs
(\ref{hydro2}--\ref{hydro3}).
 
An equation of state  $p=p(\rho,\epsilon)$ closes the system.  In this
paper   we  used   an   ideal  gas   equation   of  state   $p=(\gamma
-1)\rho\epsilon$ with $\gamma=5/3$.
 
The  hydrodynamic equations  Eqs  (\ref{hydro1}--\ref{hydro3}) can  be
rewritten in a slightly different form:
\begin{equation}
\frac{\partial  {\bf u}}{\partial t}  + \frac{\partial{\bf  f(\bf u)}}
{\partial   x}   +\frac{\partial{\bf    g(\bf   u)}}{\partial   y}   +
\frac{\partial{\bf h(\bf u)}}{\partial z}= {\bf s(\bf u)}
\label{hypersys}
\end{equation}
\noindent
where ${\bf u}$ is the vector of {\it unknowns}(conserved variables):
\begin{equation}
{\bf u} = [ \delta , m_x,m_y,m_z, E] \ \ ,
\end{equation}
\noindent
the  three {\it flux}  functions ${\bf  F}^{\alpha} \equiv  \{{\bf f},
{\bf g},{\bf h}\}$ in the spatial directions $x, y, z$, respectively ,
are defined by
\begin{eqnarray}
{\bf f(\bf u)} &=&  \left[ \frac{m_x}{a} , \frac{m_x^2}{(\delta + 1)a}
+     \frac{p}{a\rho_{_{B}}},     \right.     \nonumber\\     &&\left.
\frac{m_xm_y}{(\delta+1)a},                 \frac{m_xm_z}{(\delta+1)a},
\frac{(E+p)m_x}{a(\delta + 1)} \right]
\end{eqnarray}
\begin{eqnarray}
{\bf g(\bf u)}  &=& \left[ \frac{m_y}{a} , \frac{m_xm_y}{(\delta+1)a},
\right.   \nonumber\\   &&  \left.   \frac{m_y^2}{(\delta   +  1)a}  +
\frac{p}{a\rho_{_{B}}},                     \frac{m_ym_z}{(\delta+1)a},
\frac{(E+p)m_y}{a(\delta + 1)}\right]
\end{eqnarray}
\begin{eqnarray}
{\bf h(\bf u)}  &=& \left[ \frac{m_z}{a} , \frac{m_xm_z}{(\delta+1)a},
  \right.    \nonumber\\    &&   \left.    \frac{m_ym_z}{(\delta+1)a},
  \frac{m_z^2}{(\delta     +     1)a}    +     \frac{p}{a\rho_{_{B}}},
  \frac{(E+p)m_z}{a (\delta + 1)}\right]
\end{eqnarray}
\noindent
and the {\it sources} ${\bf s}$ are
\begin{eqnarray}
{\bf   s(\bf    u)}   &=&   \left[   0    ,-\frac{(\delta   +1   )}{a}
\frac{\partial\phi}{\partial   x}-   Hm_x   ,   \right.    \nonumber\\
&-&\frac{(\delta  +1  )}{a}  \frac{\partial\phi}{\partial y}-  Hm_y  ,
-\frac{(\delta     +1    )}{a}     \frac{\partial\phi}{\partial    z}-
Hm_z,\nonumber \\ &-& 3H(E+p)  - \frac{\rho_{_{B}}H m^2}{(\delta + 1)}
- \frac{m_x\rho_{_{B}}}{a}\frac{\partial\phi}{\partial       x}      -
\frac{m_y\rho_{_{B}}}{a}\frac{\partial\phi}{\partial  y}  \nonumber \\
&-&          \left.\frac{m_z\rho_{_{B}}}{a}\frac{\partial\phi}{\partial
z}\right]
\end{eqnarray}
where  $m_x=(\delta +1)v_x$,  $m_y=(\delta  +1)v_y$, and  $m_z=(\delta
+1)v_z$.
 
System (\ref{hypersys}) is  a three-dimensional {\it hyperbolic system
of  conservation laws}  with  sources ${\bf  s}({\bf  u})$.  From  the
numerical  point  of  view  is  important to  introduce  the  Jacobian
matrices associated to the fluxes:
\begin{equation}
{\bf  \cal  A}^{\alpha}  =  \frac{\partial{\bf  F}^{\alpha}({\bf  u})}
{\partial {\bf u}}
\label{A}
\end{equation}
\noindent
Hyperbolicity demands that any real linear combination of the Jacobian
matrices $\xi_{\alpha} {\bf \cal A}^{\alpha}$ should be diagonalizable
with real  eigenvalues (LeVeque 1992).  This is  of crucial importance
from the numerical point of view.
 
The  spectral decompositions of  the above  Jacobian matrices  in each
direction,  i.e., the  {\it  eigenvalues} and  {\it eigenvectors}  are
explicitly written in Quilis, Ib\'a\~nez \& S\'aez (1996).
 
The sources do not contain any term with differential operators acting
on hydrodynamical  variables ${\bf u}$. This is  an important property
consistent  with   the  fact   that  the  left   hand  side   of  Eq.\
(\ref{hypersys}) defines a hyperbolic system of conservation laws.

\subsection{The hydro code}

The mathematical properties resulting from the hyperbolic character of
the system of  equations (\ref{hypersys}) allow us to  design a set of
numerical  techniques known  as {\it  high-resolution shock-capturing}
(HRSC). These  techniques are  the modern implementation  of Godunov's
original method (Godunov 1959, see Laney (1998) for a modern review on
Godunov schemes and Eulerian methods).

The  HRSC  techniques  have   several  key  ingredients  such  as  the
reconstruction  procedure,  the  Riemann  solver, and  time  advancing
schemes  which can vary  in different  implementations.  Nevertheless,
all  of these  implementations share  the same  basic  properties: the
ability to handle shocks,  discontinuities and strong gradients in the
integrated quantities, and excellent conservation properties.
 
Our basic hydro solver is  based on a particular implementation of the
HRSC  methods --  see Quilis,  Ib\'a\~nez  \& S\'aez  (1996) for  more
details. The main ingredients of our solver are the following:
 
\begin{enumerate}
\item  It is  written  in {\it  conservation  form}.  This  is a  very
important  property  for  a  numerical algorithm  designed  to  solve,
numerically, a hyperbolic system of conservation laws. That is, in the
absence  of  sources, those  quantities  that  ought  to be  conserved
--according to  the differential equations-- are  exactly conserved in
the difference form.

\item  {\it  Reconstruction  procedure}.   This procedure  allows  the
method to  gain resolution by  reconstructing, through interpolations,
the  distribution of the  quantities within  the numerical  cells.  In
order   to    increase   spatial   accuracy,   we    have   used   two
cell-reconstruction   techniques.  We   have   implemented  a   linear
reconstruction  -- first  order  --, with  the  {\it minmod}  function
(Quilis, Ib\'a\~nez  \& S\'aez  1994) as a  slope limiter.   With this
reconstruction, our  algorithm is a MUSCL-version (van  Leer 1979) and
second order accurate in space.   We have also implemented a parabolic
reconstruction (PPM) subroutine according  to the procedure derived by
Colella \&  Woodward (1984).   With the parabolic  reconstruction, the
algorithm is third order accurate in space. Statements on the order of
the scheme must be taken with caution, as the mathematical proofs only
exist  for  the  one-dimensional  case,  being the  order  reduced  in
multidimensional extensions. In any case,  the higher the order of the
method  in  1D,  the  better  the  accuracy  in  the  multidimensional
applications.

Hence,  from   the  cell-averaged  quantities   ${\bf  u}_{i,j,k}$  we
construct, in each direction, a piecewise linear or parabolic function
which   preserves   monotonicity.    Thus,   the   quantities,   ${\bf
u}_{i+{1\over    2},j,k}^R,    {\bf    u}_{i,j+{1\over    2},k}^R,{\bf
u}_{i,j,k+{1\over   2}}^R$  and  ${\bf   u}_{i+{1\over  2},j,k}^L,{\bf
u}_{i,j+{1\over  2},k}^L,   {\bf  u}_{i,j,k+{1\over  2}}^L$,   can  be
computed; the  superindices $R$ and $L$  stand for the  values at both
sides of  a given interface  between neighbour cells. These  values at
each side  of a given interface  allow us to define  the local Riemann
problems. The numerical fluxes can be computed through the solution of
these local Riemann problems.
 
\item {\it Numerical fluxes} at  interfaces. We have used a linearized
Riemann solver following  an approach similar to the  one described by
Roe  (1981).  The  procedure,  applied in  each  direction, starts  by
constructing  the  corresponding  numerical  flux according  to  Roe's
prescription  and in  order to  do that  it is  necessary to  know the
spectral  decomposition of  the Jacobian  matrix  ${\cal A}^{\alpha}$.
The numerical flux associated with the $x$-direction is:
\begin{eqnarray}
{\widehat   {{\bf    f}}}_{i+{1\over   2},j,k}   =   \frac{1}{2}\left(
\frac{}{}\right.   {\bf f}({\bf  u}_{i+{1\over  2},j,k}^{L}) &+&  {\bf
f}({\bf    u}_{i+{1\over   2},j,k}^{R})    \nonumber\\    &-&   \left.
\sum_{{\eta}  =  1}^{5}  \mid \widetilde{\lambda}_{\eta}^x\mid  \Delta
\widetilde {\omega}_{\eta} {\widetilde {\bf R}}^x_{\eta} \right)
\end{eqnarray}
\noindent
where   ${\widetilde   {\lambda}}_{\eta}^x$   and  ${\widetilde   {\bf
R}}_{\eta}^x$  ($\eta=1,2,3,4,5$) are,  respectively,  the eigenvalues
and the $\eta$-right eigenvector of the Jacobian matrix:
\begin{eqnarray}
{\cal A}^x{ \rm _{i+{1\over  2},j,k}} = \left(\frac{\partial \bf f(\bf
u)  \rm} {\partial\bf  u  \rm} \right)_{{\bf  u}=( {\bf  u}_{i+{1\over
2},j,k}^{L}+ {\bf u}_{i+{1\over 2},j,k}^{R})/2}.
\end{eqnarray}

These are calculated in the  state which corresponds to the arithmetic
mean  of the states  at each  side of  the interface.   The quantities
$\Delta  \widetilde  {\omega}_{\eta}$  --   the  jumps  in  the  local
characteristic  variables through each  interface-- are  obtained from
the following relation
\begin{eqnarray}
{\bf    u}^{R}    -    {\bf    u}^{L}   =    \sum_{\eta    =    1}^{5}
\Delta\widetilde{\omega}_{\eta} {\widetilde {\bf R}}_{\eta}^x
\end{eqnarray}
\noindent
where $\widetilde {\lambda}_{\eta}^x$, $\widetilde {\bf R}_{\eta}$ and
$\Delta \widetilde  {\omega}_{\eta}$, are functions of  $\bf u$, which
are calculated at each interface and, consequently, they depend on the
particular  values  ${\bf u}^{L}$  and  ${\bf  u}^{R}$. The  numerical
fluxes in the $y$ direction, $\hat {\bf g}$, and $z$ direction , $\hat
{\bf h}$, are obtained in an analogous way.

\item {\it Advancing  in time}.  Once the numerical  fluxes $\hat {\bf
f}$,$\hat {\bf  g}$, and  $\hat {\bf h}$  are known, the  evolution of
quantities ${\bf u}_{i,j,k}$ is governed by
 
\begin{eqnarray}
\label{runge}
\frac{d{\bf u}_{i,j,k}}{d  t}&=& -\frac{\hat{\bf f}_{i+{1\over 2},j,k}
  -\hat{\bf   f}_{i-{1\over    2},j,k}}   {\Delta   x_i}   \nonumber\\
-\frac{\hat{\bf g}_{i,j+{1\over 2},k}- \hat{\bf g}_{i,j-{1\over 2},k}}
{\Delta   y_j}  &-&\frac{\hat{\bf   h}_{i,j,k+{1\over   2}}-  \hat{\bf
    h}_{i,j,k-{1\over 2}}} {\Delta z_k} + {\bf s}_{i,j,k}
\end{eqnarray}
\end{enumerate}
 
An  ordinary differential  equation (ODE)  solver derived  by  Shu and
Osher (1988)  is used to solve  Eq (\ref{runge}). It is  a third order
Runge-Kutta  that  does  not  increase  the  total  variation  of  the
numerical solution and preserves the conservation form of the scheme.

Gravity is  included in  the gas evolution  through the source  term ,
${\bf  s}_{i,j,k}$,  in Eq.   (\ref{runge}).   This  term includes the
gradient of the gravitational potential which is produced by the total
mass distribution, gas plus dark  matter.  The procedure used to solve
Poisson's equation (\ref{poisson1}) is described in Sec. 2.4.
 
The criteria to select the time  step is very important and it must be
considered globally  with other time constrains that  are unrelated to
the  gas part.   Therefore,  we  will discuss  it  in the  forthcoming
section 2.5.
 
\subsection{Dark matter dynamics} 

The dark  matter is  treated as a  collisionless system  of particles.
Each of these particles evolves obeying the following equations:
\begin{equation}
\label{dm1}
\frac{d{\bf x}}{dt}= \frac{\bf v}{a}
\end{equation}
\begin{equation}
\label{dm2}
\frac{d{\bf v}}{dt}= - \frac{\bf{\nabla \phi}}{a} - H{\bf v}
\end{equation} 
\noindent
where ${\bf  x}$, ${\bf v}=a(t)\frac{d{\bf x}}{dt}=  (v_x, v_y, v_z)$,
and $\phi(t,{\bf x})$ are, respectively, the Eulerian coordinates, the
peculiar velocity, and the peculiar Newtonian gravitational potential.

When $\phi(t,{\bf x})$  is known, the position and  velocities of each
one of the dark matter particles can be updated from the previous time 
 step.

In  our code  we solved  these equations  using a  Lax-Wendroff scheme
which is  second order. We  summarize the steps  to go from  time step
$n$, where  all the variables are  known, to the step  $n+1$, using an
intermediate step $t^{n+{1\over2}}=t^n + {{\Delta t}\over2}$:
\begin{enumerate}
\item{}Compute the intermediate step:
\begin{equation}
{\bf  x}^{n+{1\over2}}={\bf  x}^n  + {1\over2}  \frac{{\bf  v}^n}{a^n}
\Delta t
\label{dm11}
\end{equation}
\begin{equation}
{\bf   v}^{n+{1\over2}}={\bf  v}^n   -   {1\over2}\left  [\frac{\nabla
\phi^n}{a^n} + H^n{\bf v}^n\right]\Delta t
\end{equation}
\item{}Step $n+1$:
\begin{equation}
{\bf x}^{n+1}={\bf x}^n+\frac{{\bf v}^{n+{1\over2}}} {a^{n+{1\over2}}}
\Delta t
\end{equation}
\begin{equation}
{\bf  v}^{n+1}={\bf   v}^n  -  \left[\frac{\nabla  \phi^{n+{1\over2}}}
{a^{n+{1\over2}}}  + H^{n+{1\over2}}{\bf v}^{n+{1\over2}}\right]\Delta
t
\label{dm22}
\end{equation}
\noindent
where  the potential  at intermediate  step,  $\phi^{n+{1\over2}}$, is
 computed   using  a  linear   extrapolation  from   $\phi^{n-1}$  and
 $\phi^{n}$.
\end{enumerate}

In  order to  recover the  continuous  density field  for dark  matter
component, $\rho_{_{DM}}$, we use triangular shaped cloud (TSC) scheme
(Hockney \& Eastwood 1988) at each time step.
 
\subsection{Poisson solver}
The   gravitational  potential  is   computed  by   solving  Poisson's
equation. As two  components are considered, gas and  dark matter, the
source in Poisson's equation is the total density contrast:
\begin{equation}
\label{poisson}
\nabla^2\phi = \frac{3}{2}  H^2 a^2 \delta_{T}=\frac{a^2}{2} (\rho_b +
\rho_{_{DM}} - \rho_{_{B}})
\end{equation}
where         $\delta_{_{T}}=\delta_{b}+\delta_{_{DM}}+1$         when
$\delta_{b}=\rho_b/\rho_{_{B}}                 -1$                 and
$\delta_{_{DM}}=\rho_{_{DM}}/\rho_{_{B}}-1$.

Poisson's  equation  (\ref{poisson})  is  solved  using  Fast  Fourier
Transform  (FFT) methods  (Press et  al. 1996).   The FFT  is  used as
follows:
\begin{enumerate}
\item{}The density contrast in physical space --with suitable boundary
conditions-- is the starting point.
\item{}A  FFT  gives $\delta({\bf  {k}})$  (the  Fourier component  of
$\delta$).
\item{}Poisson's  equation in  Fourier space,  $\phi({\bf {k}})=G({\bf
{k}})  \delta({\bf  {k}})  $,  --  being $G({\bf  {k}})$  the  Green's
function -- is used to get $\phi({\bf {k}})$.
\item{}The inverse  FFT leads to the  required gravitational potential
in physical space.
\end{enumerate}

\subsection{The time step criteria}

In  order   to  solve  numerically   the  Eq  (\ref{runge})   and  Eqs
(\ref{dm1}--\ref{dm2}), we need to  choose a time step.  The numerical
stability  of the methods  used to  integrate these  equations imposes
several criteria  on the time  step.  At each numerical  iteration, we
compute  several   time  steps   given  by  the   different  stability
conditions.   The most  restrictive  of  all of  them  is selected  to
advance the gaseous  and dark matter components.

The time step criteria we consider are the following:
\begin{enumerate}
\item{}  Courant  time. It  is  based  on  the Courant  condition  for
stability of  an algorithm to solve partial differential 
equations. We  compute $\Delta t_C$ as:
\begin{equation}
\Delta t_C=CFL_1 \times \frac{\Delta x}{c_s+max(|v_x|,|v_y|,|v_z|)}
\end{equation}
\noindent
where $CFL_1$  is dimensionless factor between  0 and 1,  and $c_s$ is
the sound speed. This quantity is computed for all cells and the final
Courant time step is $\Delta t_C = min (\Delta t_C)_{i,j,k}$, $\forall
i,j,k$.  Typically, we use $CFL_1=0.5$.

\item{}  Dark  matter  particle  cell-crossing time.  We  compute  the
cell-crossing time  for the fastest  dark matter particle,  and define
the new time step as a fraction of this crossing time:
\begin{equation}
\Delta t_{DM} = CFL_2 \times \frac{\Delta x}{max(|{\bf v}|)}
\end{equation}
\noindent 
where we choose $CFL_2=0.2$.

\item{} Expansion  time. The equations we are  considering have source
terms  which include factors  due to  the cosmological  expansion.  At
early times  in particular, these  factors can be important  and their
time variations  introduce another  time step constraint.   We compute
this  new  time step  by  imposing  a maximum  change  of  2\% in  the
expansion  of the  Universe.  In  the  case of  flat Universe  without
cosmological constant, the time step is:
\begin{equation}
\Delta t_{e} = 0.015 t
\end{equation}

\item{} As  an extra  criteria to avoid  unexpected problems,  we also
introduce another time step ,  $\Delta t_{in}$, which prevents the new
time step to  increase more than 25\% of the global  time step for the
previous iteration.
\begin{equation}
\Delta t_{in} = \Delta t^{n-1} + 0.25 \Delta t^{n-1}
\end{equation}

\end{enumerate}

The  global time  step is  defined as  the most  stringent of  all the
previous time steps:
\begin{equation}
\Delta t = min (\Delta t_C,\Delta t_{DM},\Delta t_{e},\Delta t_{in})
\end{equation}

At the beginning  of the cosmological simulations, $\Delta  t_{e} $ is
the  dominant  time criteria  but  $\Delta  t_C$  and $\Delta  t_{DM}$
quickly take over.

\section{The Adaptive Mesh Refinement strategy}

The fundamental idea behind the  AMR technique is to overcome the lack
of resolution associated with  the fix grid Eulerian description.  The
basic idea is simple. Regions  in the original computational domain in
which improved  resolution is required are selected  according to some
criteria (see  Sec. 3.1).  These  new computational domains,  which we
call {\it  child grids} or {\it  patches}, are remapped  with a higher
number of  cells and therefore  with better resolution. The  values of
the different  quantities defined on  the child grids are  obtained by
interpolating from  the {\it parent  grid}.  Once the child  grids are
built, they can  be evolved as an independent  computational domain by
using  the  same  methods  we  have  described  in  Sec  2.   Although
conceptually   simple,  there   are  severe   technical  complications
concerning with the communication  among the different patches and the
boundary problems at different levels.
 
Our implementation of  the AMR technique follows the  one described in
Berger \& Colella(1989).

\subsection{Creating the grid hierarchy}

The first step in the construction  of the hierarchy of patches is the
coarse basic grid on which all the relevant quantities are known. From
this starting  point, some  criteria must be  applied to  decide which
cells are  {\it refinable}.  These criteria are  application dependent
and may need to be modified in certain cases.  Generally speaking, our
code uses two conditions: i)  if quantities (like density or pressure)
are larger than a given  threshold, and ii) if gradients of quantities
are steeper than a given  limit. Depending on the applications, we may
ask for only one of these conditions to be satisfied, whereas in other
cases both conditions must apply. The routine controlling this process
can easily  incorporate new conditions.  All the  numerical cells from
the coarse grid which satisfy  the refinement criteria are labelled as
{\it refinable}.

In  order  to illustrate  the  process  of  patch generation,  let  us
describe in detail  the mechanism for creating new  patches at level ,
$l+1$, once all the relevant information is known at lower level, $l$.

Let  us begin  with  the level  , $l$,  where  according to  a set  of
refinement criteria,  we have identified  all the cells  which fulfil
these conditions.  Then, we select among the {\it refinable} cells the
one which maximizes the refinement  criteria (i.e. if the criteria are
based  on  local   density,  we  chose  the  cell   with  the  highest
density). Around this maximum, the  minimal patch is created by adding
two cells  along each coordinate  direction.  Once a  patch containing
$5\times5\times5$ cells  is created, the  patch is extended  two cells
along x-axis direction  at the high-x end of the  patch, such that the
new  dimension of the  patch becomes  $7\times5\times5$ cells.   If the
number of  {\it refinable} cells  after the extension  increases, then
the extension is  accepted , otherwise this edge  remains fixed to the
value before  the extension.  The same procedure  is repeated  for the
low-x end of the patch. In  this way, the extension of the patch along
x-axis direction  is controlled.   Exactly same mechanism  is repeated
for y- and z-axes.  The patch extension procedure goes on until no new
{\it refinable}  cells are  included in the  patch, and  therefore its
dimensions remain  unchanged.  At this  point, the patch  is perfectly
defined.  All  the cells  contained in the  recently formed  patch are
removed from the list of {\it  refinable} cells at level $l$. Then, we
look again for the remaining  {\it refinable} cell which maximizes the
refinement criteria, and the process is repeated. The procedure goes on
until all  {\it refinable} cells  have been allocated in  patches, and
therefore, all patches at levels $l+1$ have been defined.

According to this  mechanism, the basic coarse domain  is divided into
patches which contain the {\it  refinable} cells.  As an extra control
to avoid the existence of patches with very few {\it refinable} cells,
we  introduce   a  new  quantity.    The  efficiency  of   the  patch,
$\varepsilon$, is defined as $\varepsilon=N_r/N_t$, where $N_r$ is the
number of  {\it refinable} cells in  the patch and $N_t$  is the total
number of cells in the same  patch.  This is a free parameter -- which
can be  adjusted depending on the  application -- and  it controls the
minimal efficiency.  Therefore, patches  with an efficiency lower than
the chosen threshold are discarded.

As  an  additional  precaution,  every  patch  is  extended  in  every
direction by adding two extra cells.

When regions on  the parent grids are already  identified and defined,
they are remapped with higher resolution by splitting the coarse cells
into new, smaller cells.  The ratio, $r$, between the coarse cell size
$\Delta x$ and the finer cell  size $\Delta x\prime$ is a free integer
parameter depending on the particular AMR implementation.  In our code
we  have chosen,  $r=\Delta  x/\Delta x\prime=2$.   This  choice is  a
compromise between gaining  resolution and avoiding possible numerical
instabilities that  can results  from using a  too high value  of $r$.
The method previously described  produces patches with a boxy geometry
and cubic cells ($\Delta x=\Delta y=\Delta z$) at any level.

At this  point, the  geometry of the  patches, their positions  in the
parent grid , and their new  number of cells, are known. The next step
must be to  define the values of the quantities  in the equations onto
the new,  finer grids. In  order to do  this, we have  implemented two
methods:   a  {\it   trilinear}   and  a   {\it  tricubic}   monotonic
interpolation, respectively (see for instance Press et al. 1996).

The  main difference  between  the two  interpolation  methods ,  i.e.
trilinear and  tricubic, is their order and  therefore their accuracy.
In  applications where  few refinement  levels are  required, tricubic
interpolation produces better results.  When the number of refinements
is high  and the size of  cells is small enough,  both methods produce
similar results,  though the tricubic interpolation  is slightly more 
CPU intensive.

The procedure described above  can be applied recursively. The patches
formed from the coarse grid, now become parent grids. The process can
be applied  to them, thus producing a  new set of patches  in a higher
level of refinement.

The method described above allows us  to create a whole set of patches
at different levels which  map our computational domain with adaptable
resolution.  The grid hierarchy is reconstructed after each time step,
once the system  has been evolved in time.  At the  new time step, the
hierarchy is rebuilt with  the procedure described above. 
Only a fraction  of cells would be brand new refined
cells, in the sense that they would map a region of the computational
domain not previously covered  at the previous
time step and with the same resolution. In this situation, 
it would  be a
great   waste  of  computational   resources  to 
completely  rebuild
the grid hierarchy by interpolating from the parent levels.
In  order  to  improve  performance,  the  cells  
at a certain level of refinement
-- at the advanced time -- which were in refined patches at the same level
at the previous time  step, are simply updated with  their time evolution
within their patch.  
Only those  cells covering  regions of the computational domain,  
which were not refined at the previous time step with the same resolution, 
are assigned with interpolated values from the respective parent patch
at a lower level.

It should be noted that  the process of grid creation described above,
produces a certain  degree of overlapping between patches  at the same
level. This situation must be kept  under control by the code as we do
not wish  to have  a scenario in  which cells  -- located at  the same
coordinates but belonging to  different patches -- could be assigned
with slightly different values of the physical quantities.

The  process  of  building the  whole  set  of  nested grids  and  the
multidimensional  interpolations needed  to assign  values to  the new
created  cells, is  very CPU  demanding.  In  our  implementation, the
hierarchy may not be rebuilt every time step.  A free parameter, which
depends on the application,  controls how frequently the hierarchy can
be modified, in the meantime, it remains unchanged.

\subsection{Gas}

The gas evolution in the AMR implementation is carried out by evolving
the different  levels starting from  the basic, coarsest  level toward
the highest, most refined level.

The  coarse basic grid  ($l=0$) is  evolved with  a time  step $\Delta
t_0$, as described  in Sec 2.2.  After that,  all quantities are known
on the coarse grid at  time $t^{n+1}=t^n+\Delta t_0$. Then, we go back
to time  $t^n$ and advance  the level $l+1\,  (l=1)$ with a  time step
$\Delta t_1$  that satisfies  $\Delta t_1 \leq  \Delta t_0$.  This time
step,  $\Delta  t_1$,  is   computed  according  to  the  prescription
described  in Sec  2.5 but  for  all the  patches at  level $l=1$.  In
certain situations, $\Delta t_1$ could be equal to $\Delta t_0$, an in
principle it  seems feasible  to save  CPU time by  using such  a time
step.  However, and  in order to maintain the order  of the method, it
is recommended that $\Delta t_{l-1}/\Delta t_{l}$ be an integer number
larger  than 1.  In  our particular  implementation, we  always ensure
that this condition is satisfied.

The patches at level $l=1$ are advanced until they reach time $t=t^n +
\Delta t_{l-1}$.   At this stage,  we say that they  {\it synchronize}
with the  previous level.  This process  is repeated in  a similar way
for  all the  levels.  Thus,  all  patches at  a given  level $l$  are
advanced until $\sum \Delta t_l=\Delta t_{l-1}$.  Then both levels are
synchronized.

In order to evolve the patches  within a given level, $l$, every patch
is extended  by two cells  on each side.   These cells are  defined by
interpolation from the parent  patch at level $l-1$. The interpolation
methods used are the same as  those described in Sec 3.1.  In the case
that $\Delta t_l  < \Delta t_{l-1}$ and more that  one step is needed,
new values at the boundary  cells -- at intermediate times between $t$
and  $t+\Delta t_{l-1}$  --  must  be defined  at  each substep.   The
quantities at  these boundary cells are defined  by interpolation from
the  parent  patch  at level  $l-1$  whose  values  at times  $t$  and
$t+\Delta t_{l-1}$  are already known.   Due to the properties  of the
algorithm, lower resolution levels evolve first, and therefore, values
for the  patch quantities  at level $l-1$  , at any  intermediate time
between  $t$  and  $t+\Delta  t_{l-1}$,  can  be  obtained  by  linear
interpolation from the values at these times.

Every time that  the evolution of the patches at  level $l$ catches up
and synchronizes with  the evolution of their parent  patch $l-1$, the
code  runs a  procedure  whose task  is  to unify  the  values of  all
overlapping cells at  different patches within the level  $l$. This is
done by  defining a  master patch  per level $l$  among the  subset of
patches which  mutually overlap.  Whenever  another patch at  the same
level overlaps  with the  master, the value  of all quantities  at the
overlapping  cells are copied  directly from  the master  cells.  This
process is also crucial to ensure good conservation properties.

\subsection{Dark matter}

The  Dark  matter  part  of  the  code  also  benefits  from  the  AMR
strategy. Again, the  basic idea is to repeat  the procedure described
in Sec 2.3 to take advantage of the preexisting set of patches. 

In  order to  solve  Poisson's equation,  the  continuous dark  matter
density is required at each patch.  The dark matter density in a patch
is obtained using the TSC mesh assignment method with the grid and cell
size  corresponding to  this  patch. All  particles  within the  patch
contribute  to the  density of  this patch.  Therefore,  particles can
contribute to  the density of  different patches at  different levels,
but are  'spread' with different cloud sizes.  This approach naturally
offers  the advantage  of not  having  to specify  a softening  length
parameter.

In order  to solve Poisson's  equation (Eq.  \ref{poisson})  at levels
$l>0$ we use a successive  overrelaxation method (SOR).  These kind of
methods  solve  Eq.(\ref{poisson}) by  discretizing  the equation  and
treating it as a linear system of equations (see Press et al. 1992).
\begin{eqnarray}
\phi_{i+1,j,k}+\phi_{i-1,j,k}+\phi_{i,j+1,k}+\phi_{i,j-1,k}+\nonumber\\
\phi_{i,j,k+1}+\phi_{i,j,k-1}-6\phi_{i,j,k}={a^2\over{2}}\delta_{i,j,k}.
\end{eqnarray}

The new potential  $\phi^{new}$ is defined by an  iterative process as
follows:
\begin{eqnarray}
\phi_{i,j,k}^{new}=\omega\phi_{i,j,k}^* + (1-\omega)\phi_{i,j,k}^{old}
\end{eqnarray}
\noindent
where
\begin{eqnarray}
\phi_{i,j,k}^*=({a^2\over{2}}\delta_{i,j,k}-
\phi_{i+1,j,k}-\phi_{i-1,j,k}\nonumber\\
-\phi_{i,j+1,k}-\phi_{i,j-1,k}- \phi_{i,j,k+1}-\phi_{i,j,k-1})/6 .
\end{eqnarray}

The overrelaxation parameter, $\omega$,  is defined in the interval $1
<  \omega  <  2$.   In  order  to  find  the  optimum  value  for  the
overrelaxation  parameter,  we  have  used  the  asymptotic  Chebyshev
acceleration procedure (see Press et al. 1992). Following this method,
the number of iterations (typically $\sim 10$) is minimized.

Once  the  potential  is  known  at  each  level,  the  positions  and
velocities   of  all   particles  can   be  updated   using   the  Eqs
(\ref{dm11}--\ref{dm22}). However,  we must use the  values of $\Delta
t_l$ and $\phi_l$ corresponding  to the highest resolution level patch
that each particle  is in.  A similar methodology  of time substepping
and  synchronization to  that  described  above for  the  gas is  also
implemented for the dark matter.

Let us
note, that the code can also be applied to dark matter only problems. 
In these cases, the patches are placed according to criteria based on 
the dark matter distribution and its features. 

\subsection{Some extra tricks}

The  above description  is based  on the  idea of  repeating  the same
numerical  scheme  for  each  patch independently.   However,  several
additional  mechanisms have  to  be  put in  place  to avoid  problems
associated  with  the  'abrupt'  change  in resolution  at  the  patch
boundaries,  such  as  the  violation of  conservation  properties  or
numerical instabilities.

\begin{figure}
\psfig{file=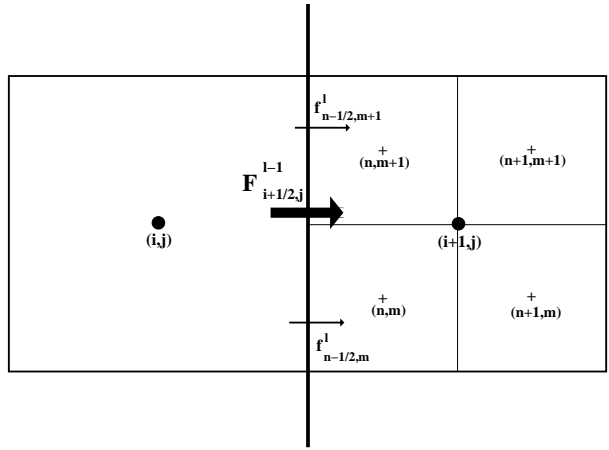,width= 8 cm}
\caption[]{The left hand boundary between a patch at level $l$ and its
parent patch at level $l-1$.}
\label{flux}
\end{figure}

From  the numerical point  of view,  boundaries represent  a difficult
issue which  deserves special attention.  Consider a  situation in two
dimensions, for  the shake of  simplicity, like the one  illustrated by
the sketch in  Figure \ref{flux}.  In this case, we  focus on the left
hand boundary, indicated  by the thick central line,  of a given patch
at level $l$.   At this point, on the left hand  side of the boundary,
we have the  cell $(i,j)$ of the parent patch at  level $l-1$ and four
cells $(n,m)$,  $(n+1,m)$, $(n,m+1)$,  $(n+1,m+1)$ on the  right side.
Underlying the four  cells of the patch at level $l$,  there is also a
parent  cell $(i+1,j)$,  which was  evolved together  with  the parent
patch at  level $l-1$.   Therefore, the values  of quantities  at cell
$u^{l-1}_{i,j}$ and $u^{l-1}_{i+1,j}$  are consistent with a numerical
flux  at their  interfaces  of $F^{l-1}_{i+1/2,j}$.  According to  our
scheme, when  the child  cells are advanced  in time  and synchronized
with their parent patch, we  substitute the value for the parent cell,
$u^{l-1}_{i+1,j}$,  by an  average of  the values  of the  child cells
$u^{l}_{n,m}$,  $u^{l}_{n+1,m}$, $u^{l}_{n,m+1}$  , $u^{l}_{n+1,m+1}$.
However,  these child cells  were updated  using two  numerical fluxes
$f^{l}_{n-1/2,m}$ and  $f^{l}_{n-1/2,m+1}$.  It  is clear that  due to
the  non  linear  character of  the  problem  and  the change  in  the
numerical    resolution   across    the   boundary,    $F^{l-1}   \neq
(f^l_{n-1/2,m}+f^l_{n-1/2,m+1})/2$.   Therefore,  when  the value  for
cell  $u^{l-1}_{i+1,j}$ is  redefined as  the average  from  its child
cells, this new value differs from the previous one which was obtained
on the coarse  grid, and is therefore linked to  its neighbours at the
same level  by a flux $F^{l-1}$(in  our example it would  refer to the
cell  $u^{l-1}_{i,j}$).   The  assignation   of  new  values  for  the
quantities at  cell $u^{l-1}_{i+1,j}$ must imply a  correction for the
values at its neighbour cells  at the level $l-1$.  These effects must
be taken  carefully into account, otherwise  severe numerical problems
can develop.

In order to overcome this problem, Berger \& Colella (1989) designed a
technique known as  {\it refluxing}. In this method,  after all values
have been advanced in time,  and patches at different levels have been
synchronized,  the  value of  the  parent  cells  next to  a  boundary
($u^{l-1}_{i,j}$ in  our example) are  modified according to  the flux
difference between the  coarse flux ($F$) and the  average of the fine
fluxes  ($f$).   In  the  above  explanation,  and  for  the  sake  of
simplicity, we have assumed that  $\Delta t_l = \Delta t_{l-1}$.  In a
general case when $\Delta t_l < \Delta t_{l-1}$, the fine fluxes ($f$)
for all the substep have to be added and compared with the coarse flux
($F$).   A detailed  description of  this  technique can  be found  in
Berger \& Colella (1989).

Another  issue related  with  the change  of  numerical resolution  at
boundaries between patches has to do with spurious oscillations of the
gravitational  forces.   The  potential  is sensitive  to  changes  of
resolution.  Therefore,  the evolution of the gas  volume elements and
dark matter particles can be affected by spurious oscillations created
by changes in the resolution of the gravitational potential across the
patch boundaries. The way we have  avoided these problem is by using a
smooth transition  method (see Anninos, Norman \&  Clarke 1994; Jessop
et al. 1994).   In our scheme, particles and gaseous  cells in a patch
at level $l$  are split in three groups: i) if  the particles or cells
are located at a distance less  than two cells (one cell for the level
$l-1$) from any  of the boundaries of the patch,  they evolve with the
potential  of the parent  patch $\phi_{l-1}$,  ii) particles  or cells
located  at  a  distance between  two  cells  and  four cells  to  the
boundaries evolve with a  linear combination of potential $\phi_{l-1}$
and  $\phi_l$, being  $\phi_{l-1}$  at  a distance  of  two cells  and
$\phi_l$ at  four cells  distance, and iii)  the rest of  particles or
cells which  evolve with  the potential $\phi_l$.  We have  found that
this method  has cured  any problem arising  from oscillations  in the
gravitational forces.

\section{Testing the code}

\subsection{Gas}

\subsubsection{Shock tube}
The so-called shock tube problems  are a set of solutions of different
Riemann problems associated with  the equations governing the dynamics
of ideal gases in 1D planar  symmetry, and when the gas is, initially,
at  rest.    They  involve,  in  general,  the   presence  of  shocks,
rarefactions and contact discontinuities. Taking into account the fact
that  the analytical solution  of the  Riemann problem  is well-known,
they are considered as standard test-beds for checking a hydro-code.

\begin{figure}
\psfig{file=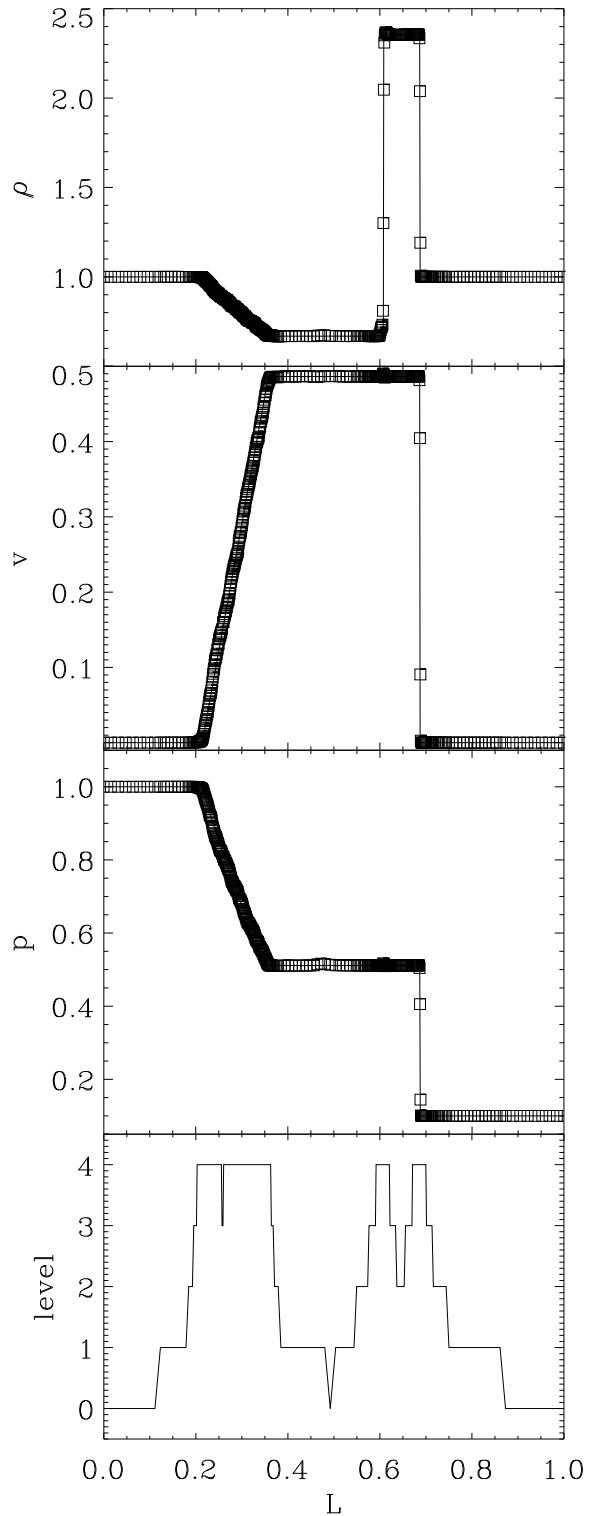,width=9cm}
\caption[]{Plot of the analytical (solid line) and numerical (squares)
solutions for  shock tube numerical  experiment.  From top  to bottom,
first, second, and third panels display density, velocity and pressure
in  arbitrary  units  ,  respectively.   The fourth  panel  shows  the
refinement  structure.  All quantities  are plotted  as a  function of
distance normalized to cube side length, L.
\label{tube}}
\end{figure}
 
In order to take advantage  of the analytical solution of the standard
shock tube problem, we  have considered a computational domain defined
by a  cube (one unit  length each edge)  in which a  discontinuity has
been placed  parallel to  one of the  cube faces.   This discontinuity
separates the  system in two  states. Initially, both states  have the
same density  ($\rho=1$) and velocity  ($v=0$).  Pressure on  the left
(right)  hand side  of  that discontinuity  is  $p=1$ ($p=0.1$).   The
coarse grid ($l=0$) used in  these computations has $64^3$ cells.  The
evolution  from the above  initial data  leads to  the formation  of a
rarefaction wave, a contact discontinuity, and a shock wave.

Figure \ref{tube} displays  the  results using  four levels  (squares)
compared with analytical results  (solid line).  Refinements have been
placed dynamically by taking  into account three conditions. The first
condition is  based on thresholds  on the relative jumps  in pressure.
For  example, for  a given  cell  $(i,j,k)$, the  relative jump  along
x-direction                is                defined                as
$\displaystyle{\frac{|p_{i+1,j,k}-p_{i-1,j,k}|}
{min(|p_{i+1,j,k}|,|p_{i-1,j,k}|)}}$.
The  jumps are  computed  for  the three  directions  (x,y,z) and  the
condition  is  satisfied if  any  of them  is  larger  than the  given
threshold.  The second condition is similar to the previous one but is
applied  to  the  density.  The  third  condition  looks  at  relative
variations of right  and left velocity derivatives.  For  a given cell
$(i,j,k)$     ,     we     define     the    right     derivate     as
$\displaystyle{D_r=\frac{v^{i+1,j,k}_x-v^{i,j,k}_x}{\Delta x}}$ and 
the left derivate
as  $\displaystyle{D_l=\frac{v^{i,j,k}_x-v^{i-1,j,k}_x}{\Delta  x}}$.   
 The  relative variation      of    the    derivate   is   defined  as
$\displaystyle{\frac{|D_r-D_l|}{min(|D_r|,|D_l|)}}$.

The first condition is specially suited for identifying shocks, second
condition finds shocks and  contact discontinuities, whereas the third
condition can track the head and tail of the rarefaction waves.
 
The code has automatically
allocated and  deallocated the  numerical patches needed  to integrate
the hydrodynamics  equations.  In  order to illustrate  the refinement
structure, the location  of each patch and their  level are plotted in
the bottom panel of Figure \ref{tube}.

The main features of the  analytical solution are recovered well.  The
shock  is  sharply  resolved  in  two or  three  cells of the 
highest level.   The  contact
discontinuity and the rarefactions  wave are also well described. Tiny
oscillations are  visible in the velocity associated  with the contact
discontinuity.

\subsubsection{Self-similar spherical collapse}

Bertschinger  (1985) presented  the solution  for the  evolution  of a
single mass perturbation in a flat Einstein-de Sitter Universe with no
cosmological constant.  Given a perturbation  of radius $R_i$  at time
$t_i$  with  overdensity  $\delta_i=\delta\rho/\rho$, shells  of  matter
surrounding  the  perturbation  start  to  decelerate  and  eventually
decouple  from the Hubble  flow at  some "turn  around" time  which is
related to the parameters defining the perturbation as follows:
\begin{equation}
r_{ta}=R_i\delta_i^{1/3}{ \left ( \frac{4t}{3\pi t_i} \right )}^{8/9}
\end{equation}

For a collisional fluid, the  infalling matter produces an increase in
pressure which  eventually results in a strong  shock wave propagating
outwards.  According  to Bertschinger's solution, the  position of the
shock  is given  by $\lambda_s=r_s/r_{ta}$.   Due to  the self-similar
character of the  problem, the solution is fully  characterized by the
dimensionless functions V, D and P:
\begin{eqnarray*}
\lambda(r,t)=\frac{r}{r_{ta}(t)}  \nonumber \\ v(r,t)=\frac{r_{ta}}{t}
V(\lambda)  \nonumber  \\  \rho(r,t)=\rho_B  D(\lambda)  \nonumber  \\
p(r,t)=\rho_{_{B}} {r_{ta}}{t} P(\lambda) .
\end{eqnarray*}

Functional forms  for V,D, and P  can be found  in Bertschinger (1985)
for different values of the adiabatic exponent $\gamma$.

We   have  set  up   an  initial   perturbation  with   $R_i=0.1$  and
$\delta_i=0.1$ at  a given time.   The simulation has been  done using
four refinement levels and a coarse grid with $64^3$ cells. This is an
extremely stringent  test for our  code, because a strong  shock moves
outwards and the self-similarity  has to be kept numerically. Moreover
, it is a  spherical problem described with a  Cartesian code. It should
be stressed that  this test also checks the  multilevel Poisson solver
described above.

The refinement criteria is based  on a Lagrangian approach which tries
to keep same mass within any cell of the simulations regardless of the
level of refinement. In order to do that, a cell in the coarse grid is
labelled  as {\it  refinable}  when  its density  is  higher than  the
background density ($\rho_{_{B}}=\frac{3H^2}{8\pi G}$); therefore, the
initial perturbation is completely refined.  The cells at first levels
are labelled as {\it refinable} when their density exceeds the initial
 mean density at the  coarse level by a  
factor of eight , that  is when the
local density  is larger than $8\rho_{_{B}}$.  The  process is similar
for  higher  levels,  thus  for  example, for  the  second  level  the
condition for a  cell to be refineable is a  local density higher than
$64\rho_{_{B}}$.

\begin{figure}
\psfig{file=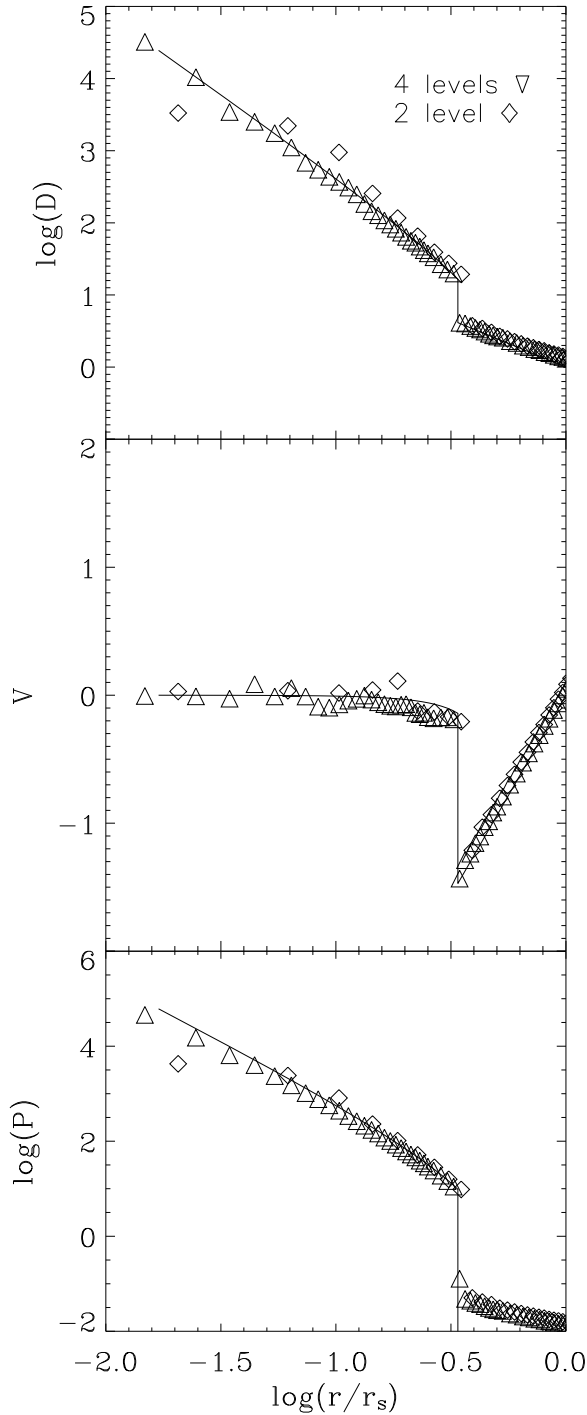,width=9cm}
\caption[]{Self-similar solution for  the evolution of a inhomogeneity
of  radius,  $R_i=0.1$,  and  density  contrast  ,$\delta_i=0.1$.  The
continuous  line represents  the analytical  solution and  the symbols
show the numerical solution (squares for the case of l=2 and triangles
for l=4).}
\label{bert}
\end{figure}

Figure  \ref{bert} shows  the  results (triangles)  compared with  the
analytical solution  from Bertschinger (1985) after  four thousand time
steps.   The numerical  solution exhibits  a good  agreement  with the
analytical  one.   It must  be  noticed  that  in the  pressure  plot,
numerical limitations  force a low,  non-zero numerical value  for the
pressure --  although with irrelevant physical consequences  -- as the
code  can  not  cope  with  a zero  value.   Self-similarity  is  well
maintained for  as long as we  have kept this test  running (more than
six thousand  time steps).  To illustrate  the convergence properties,
we display the  results (squares) when only two  refinement levels are
allowed.

\subsection{Gravity solver}

Force accuracy is  crucial to describe correctly the  dynamics of both
 the  dark  matter  and  gaseous  components. In  order  to  test  the
 properties of our gravity solver,  we present a simple test comparing
 the acceleration when a single point mass is considered.

\begin{figure}
\psfig{file=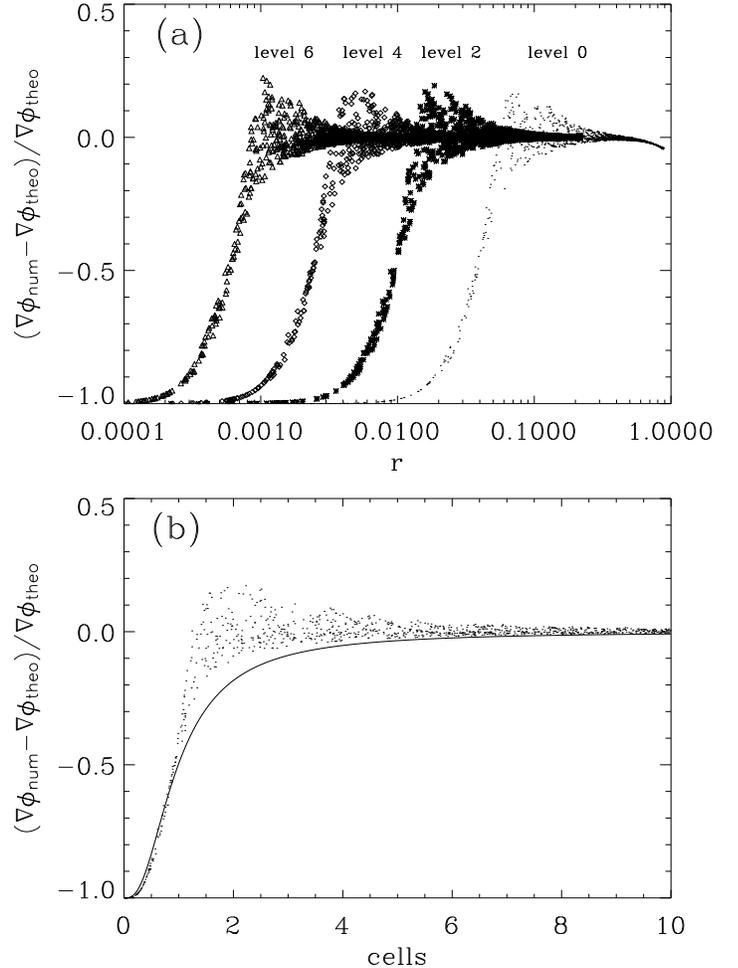,width=9cm}
\caption[]{(a) The pairwise force accuracy for the coarsest grid level
  (0),  and for  second,  fourth, and  sixth  levels, versus  particle
  separation.  (b) A comparison of the error in the force for the four
  levels  case (points) with  the error  for the  force for  a Plummer
  potential (continuous line).}
\label{gravity}
\end{figure}

We  have placed a  single point  mass ($\sim  10^{15} M_{\odot}$)  at the
centre  of  the  computational  box.  The  numerical  acceleration  is
computed for  several thousand test points randomly  placed within the
computational domain. The basic  grid has $64^3$ cells. This procedure
is repeated  with two, four, and  six nested refinements.   In each of
these  cases, the  massive particle  is always  located at  the finest
refinement.  In the upper panel  of Figure \ref{gravity}, we plot the
relative errors  of the computed accelerations when  compared with the
theoretical ones  given by $\nabla\phi=Gm/r^2$,  as a function  of the
radial distance normalized to the  distance from the box centre to the
edge  of the  computational box.   Due to  the PM-like  scheme  we are
using, the  forces are obtained  by differencing the  potential values
between neighbouring cells. When  the separation between two particles
is less  than two cells, the  forces computed with  the PM-like method
are affected  by considerable errors.   These errors are  maximal when
particles lay within the same cell, as the numerical derivative of the
potential vanishes. In Figure \ref{gravity}, this behaviour is clearly
visible when the distance approaches the cell size for each level. The
higher the level  of refinement, the smaller the  spatial range of the
force  errors,  as they  are  always  linked  to particle  separations
smaller than two cells, and the cell sizes shrink for higher levels.

Lower panel in Figure \ref{gravity}  shows the relative errors for the
numerical acceleration for the four  level case (points) and the error
for     the     acceleration      for     a     Plummer     potential,
$\phi=-Gm/(r^2+\epsilon^2)^{1/2}$.  The  softening, $\epsilon$, is set
to  the  cell size  of  the  refinement  level including  the  massive
particle. In  this case,  distance is  in units of  grid cells  of the
finest refinement level.  It should be noted that ,whereas the Plummer
acceleration marginally catches  the $1/r^2$ law at 6  or 7 cells, our
result oscillates around this value at two cells.

The  implementation of  the gravity  solver described  above naturally
fixes the force  softening to the local properties  of the solution in
an  unambiguous  way.  In  this  sense, numerical  problems  like  the
two-body  effects can  be  suppressed  because the  cell  size can  be
adapted  depending  on  the  local  density.   This  is  an  important
advantage respect to the codes using a fix softening.

\subsection{Cosmological simulation: the
Santa Barbara cluster}

The best test  we can think of for a cosmological  code in a realistic
setting  is to  compare with  the results  presented in  Frenk  et al.
(1999,  hereafter  F99).  In  this reference,  the  authors  performed
adiabatic  simulations of a  galaxy cluster  with several  codes. This
work  not only  allows a  comparison between  the different  codes and
schemes, but also establishes a standard test.

We have  adopted the initial conditions  used in F99  to simulate with
our  code  the so-called  Santa  Barbara  cluster  (SB).  The  initial
conditions were generated in order  to obtain a galaxy cluster of mass
$\sim 10^{15} M_{\odot}$ located at  the centre of a computational box
of side length  64 Mpc.  The coarse grid (l=0)  has $64^3$ cells.  The
first refinement  (l=1) is fixed covering  the central 32  Mpc and has
also  $64^3$   cells  which  are   half  the  size  of   their  parent
cells. Higher refinements are carried out according to a local density
criteria.   Two groups  of  dark matter  particles  -- with  different
masses -- are used.  At the initial time, the first group of particles
is  located within  the region  of  computational domain  that is  not
covered by the  refinement, whereas the second group  is placed on the
refinement.   Therefore, on  the part  of  the coarse  grid (l=0)  not
refined,   we   place   $32^3$   particles   with   individual   mass,
$m_{_{DM}}^{l=0} \sim  6.24 \times  10^{10} \, M_{\odot}$,  whereas on
the  first refinement  (l=1), there  are $64^3$  particles  with mass,
$m_{_{DM}}^{l=1} = m_{_{DM}}^{l=0}/8 \sim 7.8\times10^9 \, M_{\odot}$.
The cosmological  parameters assumed are  $\Omega=1, \ \ H_o=50  \ km\
s^{-1}\  Mpc^{-1}$, and  the  baryon content  is $\Omega_b=0.1$.   The
simulation was started at $z=30$.

\begin{figure*}
\psfig{file=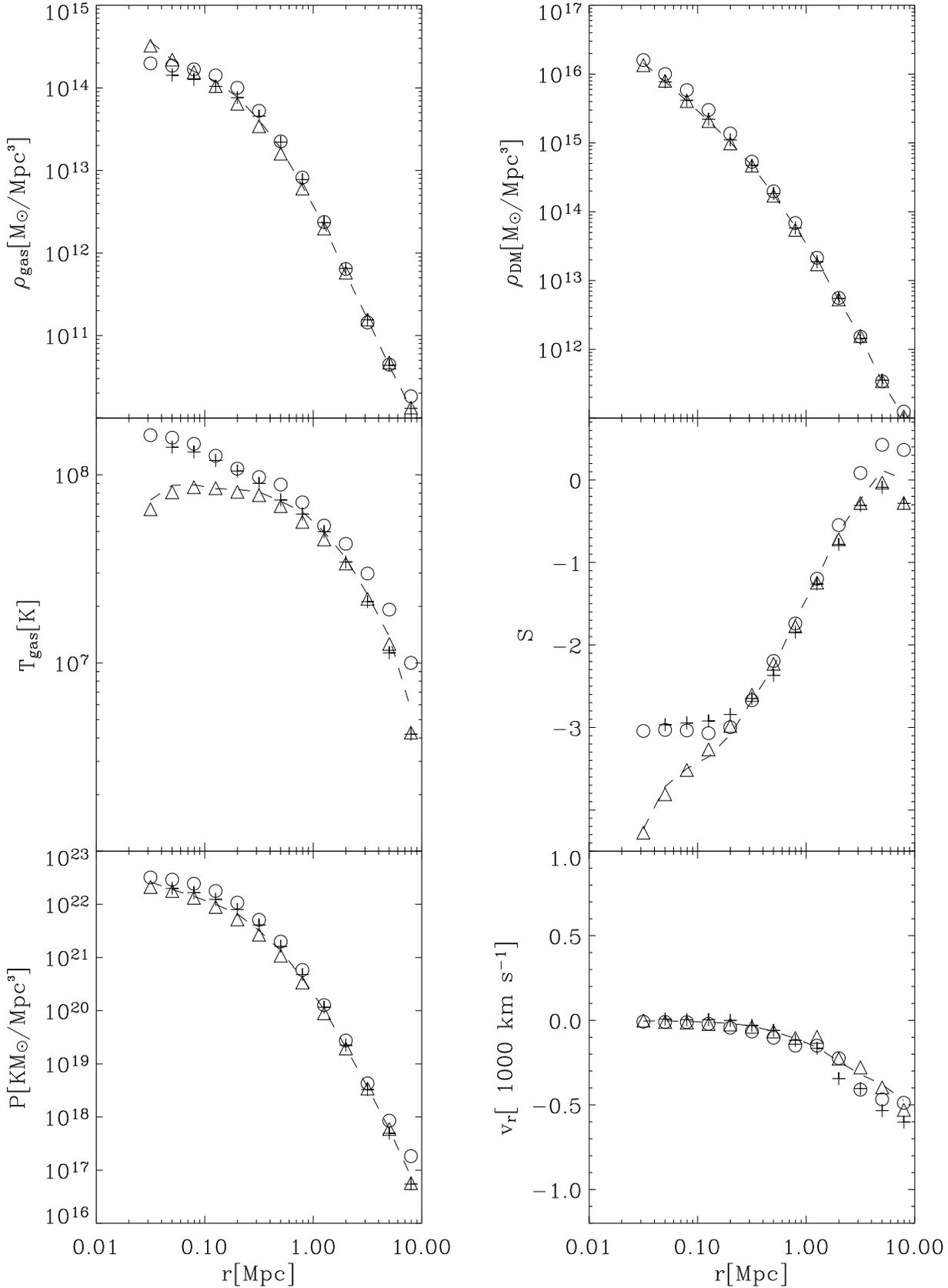,width=17 cm}
\caption[]{Radial  profiles for Santa  Barbara cluster.   Dashed lines
shows the  average results  of all the  simulations displayed  in F99.
Triangles and crosses  show the results of simulations  carried out by
Jenkins (SPH) and Bryan (SAMR), respectively. Open circles display our
results.
\label{profiles}}
\end{figure*}

The refinement criteria  is based on the local  gas density.  Any cell
with a baryonic mass larger than $3.5\times10^9M_{\odot}$ is labelled as
{\it refinable}.  This  procedure is repeated up to  six levels (l=6),
which means  that the highest  comoving spatial resolution  is $\Delta
x_{_{l=6}}=15.6\ Kpc$, although the real resolution is a factor of two
worse than this value, as the gravity part of the code needs two cells
in order to compute the gravitational force.  

In Figure \ref{profiles}
radial  profiles for  gas density,  dark matter  density, temperature,
entropy  ($S=ln[T\rho_{_{gas}}^{-(\gamma-1)}]$),   pressure,  and  gas
radial  velocity  are shown,  respectively.   The  profiles have  been
computed by  averaging the quantities  in spherical shells of  0.2 dex
logarithmic width.  The inner radius of  the first shells  is $10 kpc$
and the outer radius of the last one is $10 Mpc$.
In order to compare with  F99 results, each panel displays the average
results  of all simulations  (dashed line),  results from  Jenkins' \&
Pearce's simulation using Hydra SPH (triangles), and the simulation of
Bryan \& Norman using the  SAMR Eulerian code (crosses).  Open circles
show  the results  of  our  simulation.  Our  results  show very  good
agreement with the general trends  shown by the average results of all
simulations. There is a particularly good match with the profiles from
Bryan's SAMR code.

Nevertheless, there  are differences between  the profiles in  F99 and
the  ones we  have obtained  in our  simulation which  deserve further
comments.  Our results exhibit  a slightly steeper dark matter density
profile in the inner most regions.
In the  SB cluster simulation, there  is a merger event  at $z\sim 0$.
We would suggest that this fact could produce the small differences in
the final state of the  cluster, leading to minor discrepancies in the
comparison.  We note that our simulations at $z\sim0.09$ show a better
agreement -- in the global  quantities, radial profiles, and 2D images
-- with the results in F99.
In  any case,  we believe  our results  are consistent  with  F99.  

In addition  to the  differences previously  mentioned, 
 there  is  also a
systematic natural dispersion due to factors like the way in which the
initial  conditions  are implemented,  the  different approaches  when
averaging   to  compute   the  radial   profiles,  and   the  inherent
peculiarities of  each particular numerical algorithm.   It is notable
that similar results to ours have been obtained by Kravtsov, Klypin \&
Hoffman  (2002) when  they tested  their new  AMR-like code  using the
Santa  Barbara test.   These  authors claim  that  their profiles  are
consistent  with the  recent  results --  using  very high  resolution
simulations  -- presented  by Ghinga  et  al.  (2000),  Klypin et  al.
(2001), and Power et al. (2003).

Returning  to the  physical meaning  of our  simulations,  our results
reinforce the difference  -- first revealed in F99  by Bryan's results
-- between  SPH and  Eulerian codes.   These alarming  differences are
particularly dramatic in the thermodynamics of the gas, and quantities
like temperature  or entropy clearly manifest  different trends. There
is no doubt that these  differences are evidence of different physical
properties, and could therefore be crucial in the process of structure
formation.  New formulations of SPH, like the {\it entropy-conserving}
(Springel  \& Hernquist 2002)  which is  written in  conservative form
similar  to  the  Eulerian  schemes,  have  dramatically  improve  the
comparison. In a  detailed study, Ascasibar et al.  (2003) simulated a
cluster of galaxies -- similar to the SB cluster -- using the Eulerian
code   ART(Kravtsov,   Klypin   \&   Hoffman  2002)   and   SPH   code
GADGET(Springel \&  Hernquist 2002).  The outcome  of their simulation
showed how the results of the new SPH implementation are closer to
the ones  produced by the  Eulerian AMR codes.   Although encouraging,
there are still significant differences in the entropy and temperature
profiles.  A good example of  these differences is given by Okamoto et
al.  (2003),  where the  authors  proved  that  standard SPH  approach
produces artificial effects which can lead to unrealistic results when
studying galaxy  formation. These  spurious effects are  not necessary
linked to the artificial viscosity used by the SPH method.

\begin{figure}
\psfig{file=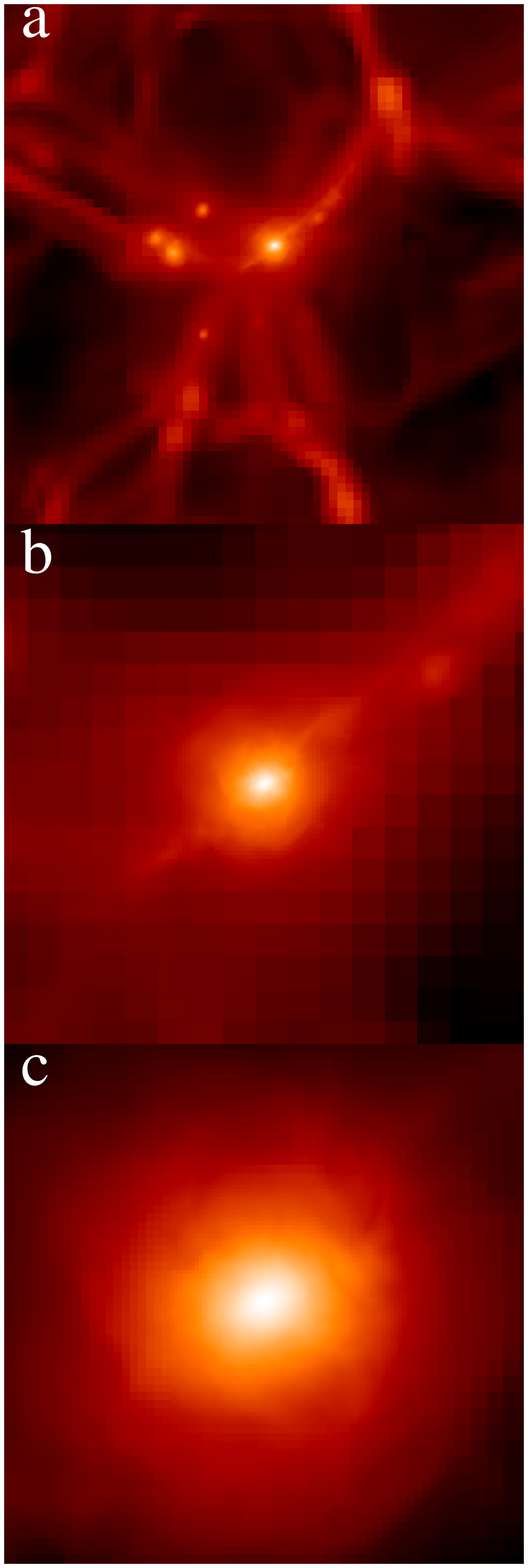,width=7 cm}
\caption[]{Logarithm  of gas  column  density along  z-axis for  three
centred regions at  redshift z=0. From top to bottom:  (a) 64 Mpc side
image of  the gas column density  integrate along the  central 32 Mpc,
(b) 16  Mpc side image of  the gas column density  integrate along the
central 8  Mpc, and  (c) 4 Mpc  side image  of the gas  column density
integrate along the  central 2 Mpc.  The 'pixelization'  effect due to
the variation  in the cell sizes  for the different  levels is clearly
visible.}
\label{panel1}
\end{figure}

Figure \ref{panel1} shows, at the final redshift $z=0$, the gas column
density along the z-axis in  the different regions.  The top panel (a)
shows a  region of 64  Mpc side length,  where the column  density has
been computed by integrating along  the central 32 Mpc. Panels (b) and
(c),  show the  column density  maps for  two zooms  into  the central
region. The images show regions of 16 and 4 Mpc side, integrated along
8 and  2 Mpc, respectively.   The 'pixelization' effect  introduced by
the variation in  the cell size corresponding to  the different levels
is  clearly visible,  as is  a web  of gaseous  filaments  linking the
structures.  It should  be noted that there are  no unphysical effects
associated  with  filaments  crossing  the boundaries  of  patches  at
different levels.

\begin{figure}
\psfig{file=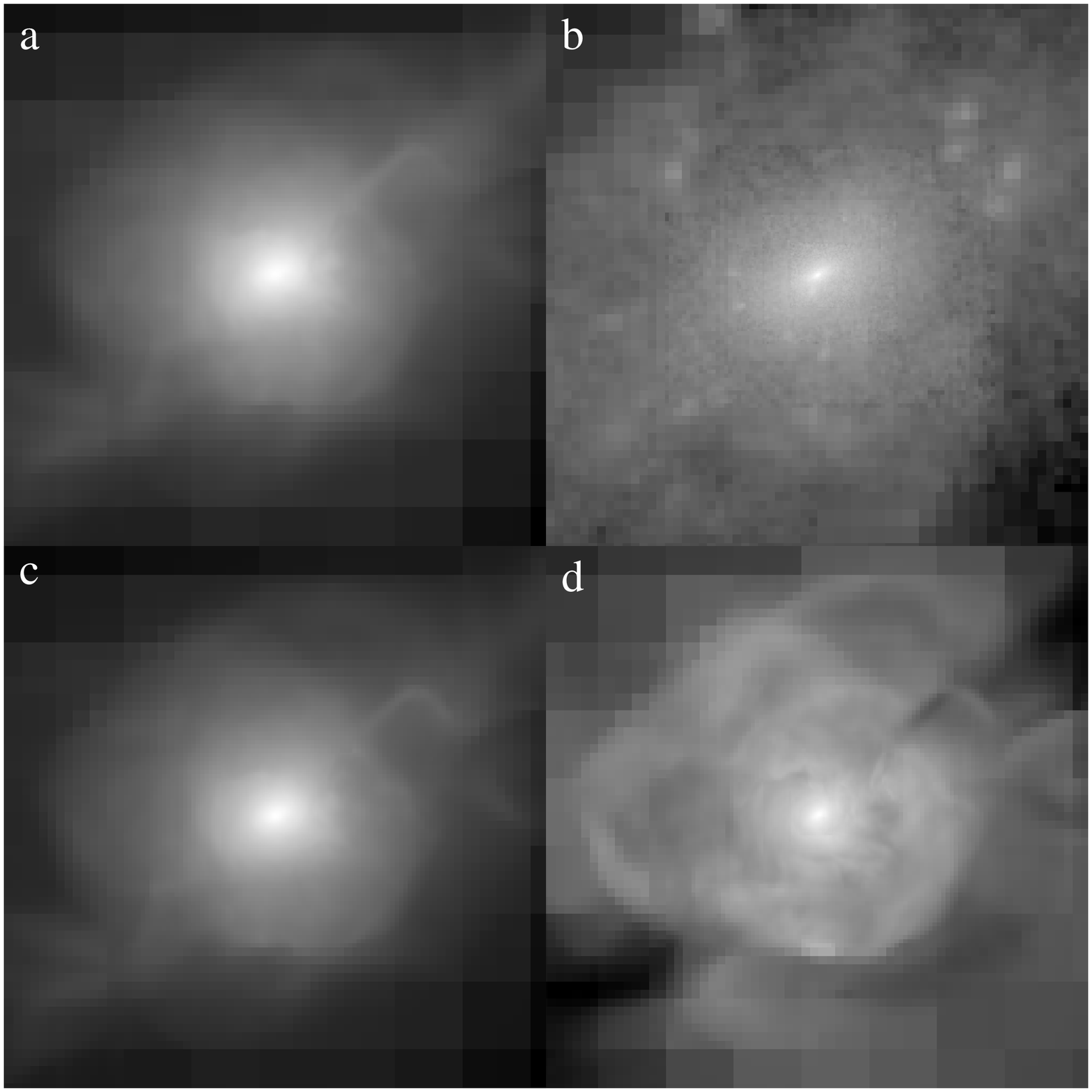,width=9 cm}
\caption[]{Logarithm  of gas  column density  (a), dark  matter column
density  (b),  X-ray  surface  brightness (c),  and  emission-weighted
temperature(d) along z-axis  for the central 8 Mpc  at z=0.  The image
sizes are $8\times8$  Mpc.  As these panels show  raw results from the
simulations, the 'pixelization' effect stands out clearly.}
\label{multipanel}
\end{figure}

In Figure \ref{multipanel}, we  present several panels displaying maps
of the  following quantities:  a) gas column  density, b)  dark matter
column density  , c) X-ray surface brightness  calculated as $\int{L_X
dl}$ where $L_X=\rho^2  T^{1/2}$, and d) emission-weighted temperature
calculated as $\int{L_X T dl}  / \int{L_X dl}$.  All images correspond
to the  projection of these quantities  along the z-axis in  the 8 Mpc
central box.   The gross  features are similar  to the images  in F99,
although,  as the average  profiles showed,  there are  differences in
some of the details. The  dark matter (Figure \ref{multipanel}, panel b)
column  density  looks  very  similar  to  results  from  the  highest
resolution  simulations  in F99.   Smoothing  and presentation  effect
apart,  the substructure in  our simulation  matches the  results from
Jenkins \& Pearce (SPH) and Bryan \& Norman (AMR).

\begin{figure}
\psfig{file=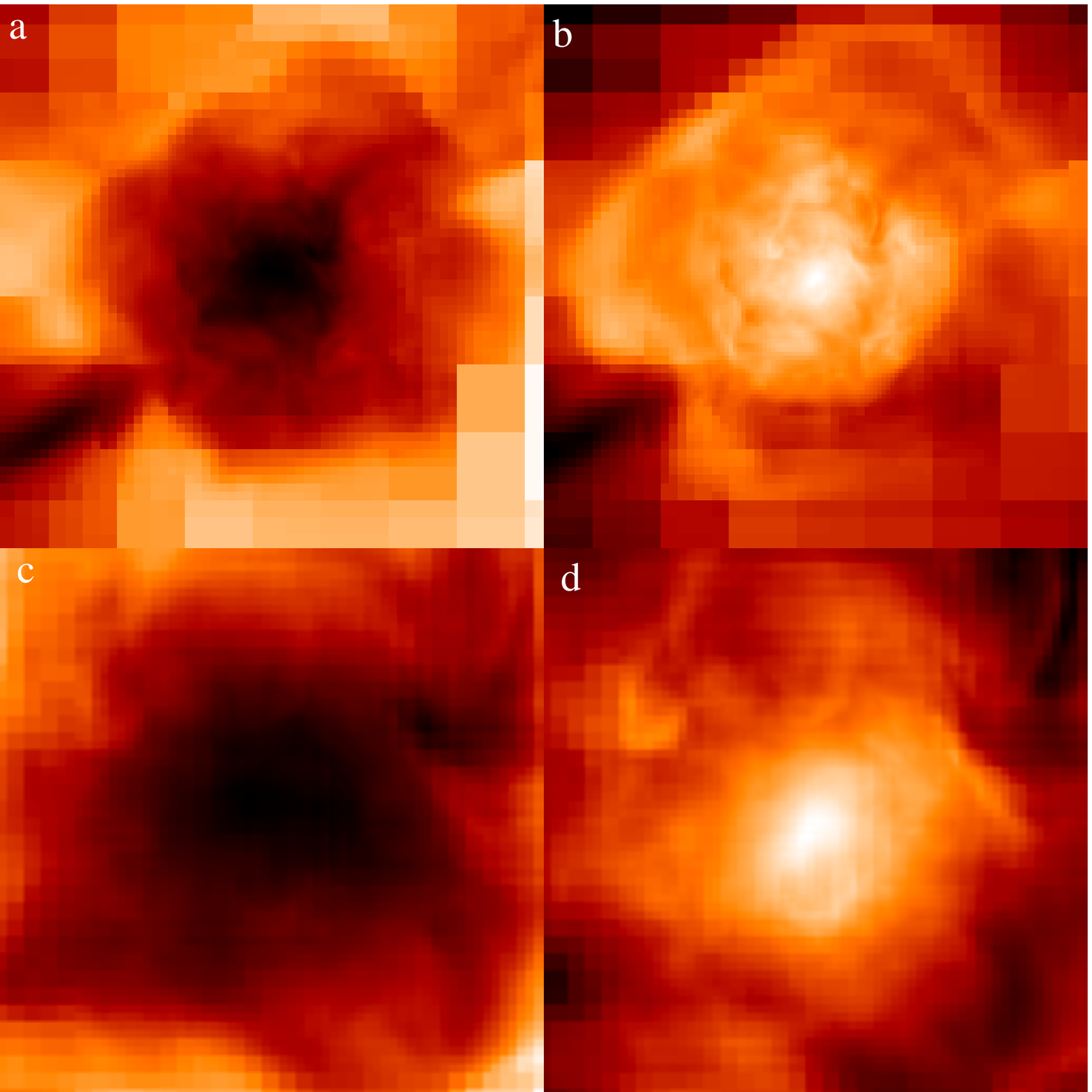,width=9 cm}
\caption[]{The structure of the shocks and turbulence in the our  
simulation of the Santa Barbara cluster. Panels  (a) and (c)
display the entropy in a very thin slice for two zoom-ins at 8 (a) and
2  (c) Mpc.  Identically  panels (b)  and  (d) show  logarithm of  the
temperature in the same regions.}
\label{multipanel2}
\end{figure}

The rich structure  of shocks, voids, gradients and  turbulence can be
shown by  displaying the entropy  and temperature within a  thin slice
containing  the  centre  of  the  cluster.   Figure  \ref{multipanel2}
displays the  entropy and  the logarithm of  temperature in  four thin
slices.  Upper (lower) panels  of Figure \ref{multipanel2} show a zoom
of the  central 8  Mpc (2 Mpc).   The images  in black and  white, are
encoded  in such  a  way that  black  (white) stands  for the  minimum
(maximum) value of the quantity plotted.

The entropy plots show a constant  entropy core (panels (a) and (c) in
Figure  \ref{multipanel2}). In  the temperature  maps (panels  (b) and
(d)) several  features are  noticeable. A rich  structure of  shock is
visible at all scales.  The  existence of cold regions embedded in the
central hot structure is another  remarkable feature.  In all cases, a
highly turbulent structure is clearly visible.  All these features can
be  studied  due  to  the  good  shock  capturing  properties  of  the
algorithm.

The Santa Barbara run has been performed using the parallel version of
the  code,  which has  been  tested  in  different architectures.   In
particular,  the run  studied  in  this section  was  performed in  an
Origin 3800 using 8 processors, and  took 130 hours of CPU time, with
a memory requirement of around 2 Gb.

\section{Conclusion}

We  have  described  and  tested  a new  cosmological  code  specially
designed  to  study  the   formation  and  evolution  of  cosmological
structures, i.e.  galaxy cluster,  galaxies, etc.  The code can follow
the  hydrodynamics of  the gas  in  a cosmological  context, the  dark
matter component, and  the gravity of the two  coupled components. The
main  ingredients  of  the  code  are:  i) a  hydro  solver  based  on
high-resolution shock-capturing techniques,  ii) a N-body solver based
on  the generalization  of the  Particle Mesh  scheme, iii)  a Poisson
solver which combines FFT and SOR techniques, and iv) an Adaptive Mesh
Refinement (AMR) scheme which allows  us to gain resolution in schemes
(i)-(iii)  by  refining the  grid  as  required  according to  several
application dependent criteria.

A set of tests, including a realistic cosmological simulation known as
the Santa  Barbara cluster, have been  performed so as  to explore the
reliability of  the code. All  of them have been  passed successfully.
The whole  package of the different ingredients  represents a powerful
code which will  be able to tackle -- with very  high spatial and time
resolution --  all sort  of challenging cosmological  problems without
giving up  an accurate description  of the physical properties  of 
the gaseous 
and dark matter components.

Previous results  using AMR simulations --  with resolution comparable
to those of  SPH -- showed substantial differences  in quantities like
temperature  and entropy  with respect  to  SPH results.   This is  an
important conclusion  -- confirmed by our simulations  -- which points
out crucial differences between the  SPH and the Eulerian methods when
describing the physics of structure formation.

AMR codes,  like the  one presented in  this paper, will  be essential
tools in the future  of numerical cosmology. They combine successfully
the best properties of previous traditional approaches: i.e. excellent
spatial resolution of the  SPH techniques, and an accurate description
of  the   hydrodynamics  of  the  problem  (handling   of  shocks  and
discontinuities, treatment  of void regions,  accurate thermodynamics,
turbulence, and  some others) inherited from the  Eulerian codes based
on Riemann solvers.

In the  near future some improvements  to the code  will be performed.
Already  as  ongoing projects,  we  are  working  to incorporate  star
formation recipes, chemical evolution, cooling, and magnetic fields.

{\bf Acknowledgements}.  VQ is a  Ram\'on y Cajal Fellow of the Spanish
Ministry  of Science and  Technology.  This  works has  been partially
supported by grant AYA2003-08739-C02-02 (partially financed with FEDER
funds).  The author  gratefully acknowledges the enlightening comments
of  the  referee  and  wishes   to  thank  the  following  for  useful
discussions and comments: R.  Bower, C.  Baugh, C.S.  Frenk, L.  Heck,
J.M$^{\underline{\mbox{a}}}$.       Ib\'a\~nez,      A.       Jenkins,
J.M$^{\underline{\mbox{a}}}$.  Mart\'{\i}, B.   Moore, and T. Okamoto.
Simulations were carried out as  part of the Virgo Consortium on COSMA
at   Institute  for   Computational   Cosmology  and   COSMOS  at   UK
Computational Cosmology Consortium,  and in the Servei d'Inform\'atica
de la Universitat de Val\`encia (CERCA and CESAR).

\end{document}